\pdfoutput=1
\documentclass[journal=jctcce,layout=twocolumn, manuscript=article]{achemso}
\usepackage[version=3]{mhchem} 
\usepackage{amsmath,amssymb}

\newcommand{\onlinecite}[1]{\hspace{-1 ex} \nocite{#1}\citenum{#1}} 

\author{Diego Prada-Gracia}
\email{diego.prada@frias.uni-freiburg.de}
\author{Francesco Rao}
\email{francesco.rao@frias.uni-freiburg.de}
\affiliation[FRIAS]
{School of Soft Matter, Freiburg Institute for Advanced Studies, Freiburg, Germany}

\title[aaa]
  {Is ion channel selectivity mediated by confined water?}

\begin{document}

\begin{abstract}

Ion channels form pores across the lipid bilayer, selectively allowing
inorganic ions to cross the membrane down their electrochemical
gradient.  While the study of ion desolvation free-energies have
attracted much attention, the role of water inside the pore is less
clear.  Here, molecular dynamics simulations of a reduced model of the
KcsA selectivity filter indicate that the equilibrium position of
Na$^{+}$, but not of K$^{+}$, is strongly influenced by confined
water.  The latter forms a stable complex with Na$^{+}$, moving the
equilibrium position of the ion to the plane of the backbone
carbonyls.  Almost at the centre of the binding site, the water
molecule is trapped by favorable electrostatic interactions and
backbone hydrogen-bonds. In the absence of confined water the
equilibrium position of both Na$^+$ and K$^+$ is identical. Our
observations strongly suggest a previously unnoticed active role of
confined water in the selectivity mechanism of ion channels.
\end{abstract}

\section{Introduction}

Neurons enable us to think, act and remember \cite{Kandel2000Principles}. At
the fundamental level, an important role is played by ion channels.  Forming
potassium-selective pores that span the cell membrane, potassium channels are
the most widely distributed channels in nature \cite{Kandel2000Principles,
Hille2001Potassium}. The breakthrough in the structure determination came from
the identification of the bacterial homolog from \emph{Streptomyces lividans}
(KcsA) \cite{Doyle1998Structure}. This channel is characterized by a tetrameric
structure in which four identical protein subunits associate around a central
ion conducting pore. At the extracellular side, the selectivity filter is
formed by a highly conserved sequence of five residues (TVGYG)
\cite{Heginbotham1994Mutations,Doyle1998Structure}.

\begin{figure}[!t]
  \begin{center}
  \includegraphics[width=80.0mm]{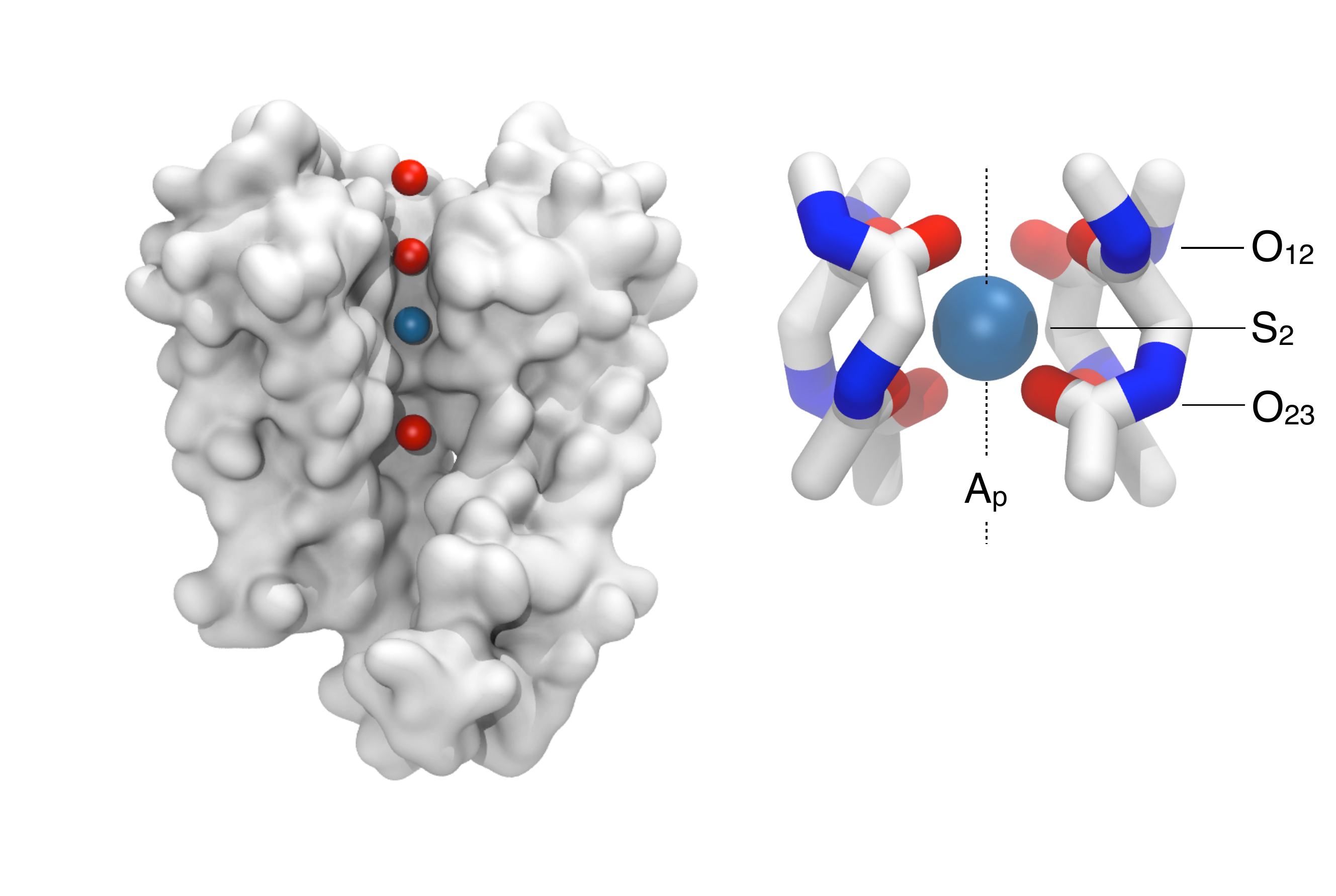}
  \caption{ {\bf The KcsA channel.} ({\em Left}) The whole protein
    channel. K$^+$ ions are depicted in color (blue for the one in the
    S2 binding site).  ({\em Right}) The reduced model of the S2
    binding site, four di-glycines harmonically constrained to the
    crystal structure are used (see Methods for details).  The
    experimental position of K$^+$ at the center of the binding site,
    the extra and intra-cellular carbonyls are labeled as $S_2$,
    $O_{12}$ and $O_{23}$, respectively.}
  \label{fig:protein_cg}
  \end{center}
\end{figure}

Originally, the high complementary of the selectivity filter to K$^+$ was
thought to be the reason for the selectivity \cite{Zhou2001Chemistry}. But it
is now clear that the interplay between structure and dynamics of the
selectivity filter is at the origin of the mechanism
\cite{Fowler2008Selectivity,roux2011,dixit2011,varma2011}. Recently, atomic
models of the selectivity filter were used to elucidate the role of 
dynamics in the process
\cite{Noskov2004Control,Asthagiri2006Role,Yu2010Two,kast2011}.  In one of these
models, the filter is reduced to the most selective binding site of KcsA,
called S2 \cite{Berneche2001Energetics}. This is done by harmonically
constraining the four backbone segments defining S2 to the experimental
conformation \cite{Noskov2004Control,Yu2010Two}. Solvation free-energy calculations
illustrated that the model is selective, allowing a statistical mechanics
treatment of the limiting cases of rigid and very flexible binding sites
\cite{Yu2010Two}.  Moreover, calculations on larger models of the filter showed that
backbone fluctuations are influenced by the presence of Na$^{+}$ or K$^{+}$ at
position S2 \cite{Asthagiri2006Role}.  These results support the view that
several aspects of selectivity can be elucidated by analyzing ion binding to 
simplified models of the filter.

Another important player in selectivity is water.  Simulation results
showed that confined water appears together with cations in the conduction pore
\cite{berneche2000molecular,guidoni2000water,domene2003potassium}.
Notwithstanding, it is not clear yet if water actively
mediates selectivity or not. 

Here, the role of confined water is investigated by molecular dynamics
simulations of a S2 binding site model, providing evidence that the
equilibrium position of Na$^+$ within the binding site is displaced by
the presence of a water molecule. Our calculations are in agreement
with a recent crystallographic study \cite{Thompson2009Mechanism} and
multi-ion free-energy calculations
\cite{Thompson2009Mechanism,Kim2011selective}. These concepts support
the idea that KcsA can bind both Na$^+$ and K$^+$ with similar
strength but different mechanism.

\section{Methods}
\noindent{\bf The S2 model.}
A reduced model of the S2 binding site of the KcsA channel (PDB code:
1K4C, \ref{fig:protein_cg}) was built with four diglycine peptides as
done in Ref.~\onlinecite{Yu2010Two}.  The heavy atoms of the reduced
model were constrained with an harmonic potential of force constant
$k=1000$ kJ/mol/nm$^{2}$ ($\approx 2.4$
$\mathrm{kcal/mol/}$\AA$\mathrm{^{2}}$).  The coordinates were
translated with the vector $(-20,0,0)$, taking the axial coordinate
$A_p$ parallel to (1,0,0). A pdb file of the reduced model is provided
as Supplementary Information (\emph{SI}).

\vspace{1mm} \noindent{\bf Molecular dynamics simulations.}
All calculations were performed with the GROMACS program
\cite{VanDerSpoel2005GROMACS,Hess2008GROMACS} and the AMBER-03 force
field \cite{Duan2003Pointcharge,Sorin2005Exploring}. A cubic box of
initial length of 4 nm solvated with TIP4P-Ew water was used
\cite{Horn2004Development}.  Simulations were integrated with the
Langevin equations ($\tau=~0.2$) at 300~K coupled with a Berendsen
barostat ($\tau_p=~1.0$~ps) \cite{Berendsen1984}.  Long range
electrostatics was computed with PME \cite{Darden1993} with a 1.0 nm
cut-off for all non-bonded interactions.  After 10 ns of
equilibration, each ion was simulated by a 100~ns long trajectory. For
both Na$^+$ and K$^+$, the starting configuration was taken as the
center of the binding site ($S_2$, see \ref{fig:protein_cg}).  To
check that there was no influence on the starting position, 20 runs of
5~ns each were further performed (Figure S1 in \emph{SI}).
Calculations performed with a TIP3P water model are in agreement with
the present analysis (see Figure S2 in \emph{SI}).

\vspace{1mm} \noindent{\bf Potential of mean force}
The potential of mean force for Na$^{+}$ and K$^{+}$ was computed with
GROMACS \cite{VanDerSpoel2005GROMACS,Hess2008GROMACS} along the pore
axis from $A_p=-5.0$ (the bulk) to $A_p=0.1$ (position
$S_2$). Umbrella sampling calculations were spaced by 0.1 \AA\ along
the axial coordinate $A_p$ with the ion restrained in the normal plane
with an harmonic potential ($k_{y}= k_{z}$=1000 kJ/mol/nm$^{2}$). An
additional harmonic potential of force constant $k_{x}=10000$ and
20000 kJ/mol/nm$^{2}$ was applied in the $-5.0<x<-2.1$ and
$-2.0<x<-0.1$ range, respectively.  After 1~ns of equilibration,
trajectories were run for 1~ns. The weighted histogram method was used
to reconstruct the potential of mean force \cite{kumar1992weighted}.

\vspace{1mm} \noindent{\bf Ion interaction energy in vacuo.}
To calculate the ion interaction energy $E_{ion}$, the cation was
harmonically restrained ($(k_{x},k_{y},k_{z})=(50000,1000,1000)$
kJ/mol/nm$^{2}$) at 0.25 \AA\ spaced positions along the axial
coordinate $A_p$ from -5.0 to 0.0 \AA. After 1~ns of equilibration,
each run was performed for 10~ns at 300 K and constant volume.

\section{Results}

The KcsA channel is shown in \ref{fig:protein_cg}.  In our simulation study, a
reduced model of the protein selectivity filter was used.  The S2 binding site
was modeled by four peptides constrained to the experimental structure (right
panel, see Methods for details) \cite{Noskov2004Control,Yu2010Two}.  Within
this model, the pore axis centered along the channel is labeled as $A_p$, while
the origin of the axis is taken as the experimental position of the K$^+$ ion.
This position is conventionally called $S_2$.  Similarly, the positions of the
carbonyl oxygens defining the binding site at the extra and intra-cellular
sides of the model are denoted as $O_{12}$ and $O_{23}$, respectively
(\ref{fig:protein_cg}).

\begin{figure}[!t]
  \begin{center}
  \includegraphics[width=80.0mm]{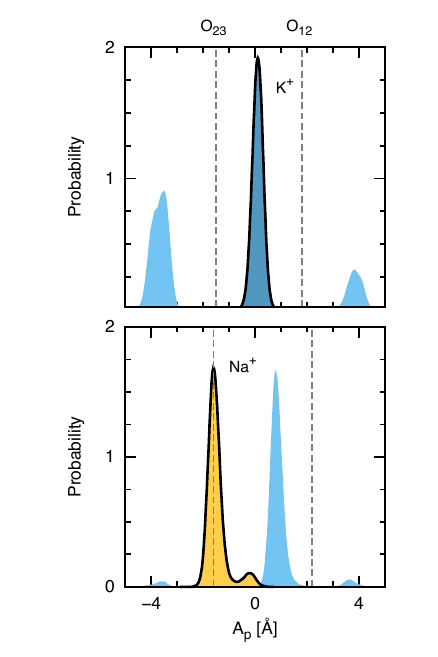}
  \caption{{\bf Probability density functions along the pore axial
    coordinate $A_p$.}  (\emph{Top}) K$^+$ (black countoured blue area);
    (\emph{Bottom}) Na$^+$ (black countoured orange area).  Position of the
    closest water molecule and of $O_{12}$ and $O_{23}$ carbonyls are
    showed as blue areas and dashed lines, respectively.}
  \label{fig:histo_1D}
  \end{center}
\end{figure}

The behavior of K$^+$ and Na$^+$ inside the S2 binding site was studied by
molecular dynamics simulations in explicit water (see Methods for details). In
both cases, the starting position of the ion was $S_2$.  \ref{fig:histo_1D}
shows the probability distribution of the ion position on the $A_p$ axis. As
expected, K$^+$ was found at position $S_2$ (black countoured blue area)
between $O_{12}$ and $O_{23}$ (dashed lines), coordinating with the eight
carbonyl oxygens of the binding site
\cite{Berneche2001Energetics,Noskov2004Control}. 

\begin{figure}[!t]
  \begin{center}
  \includegraphics[width=80.0mm]{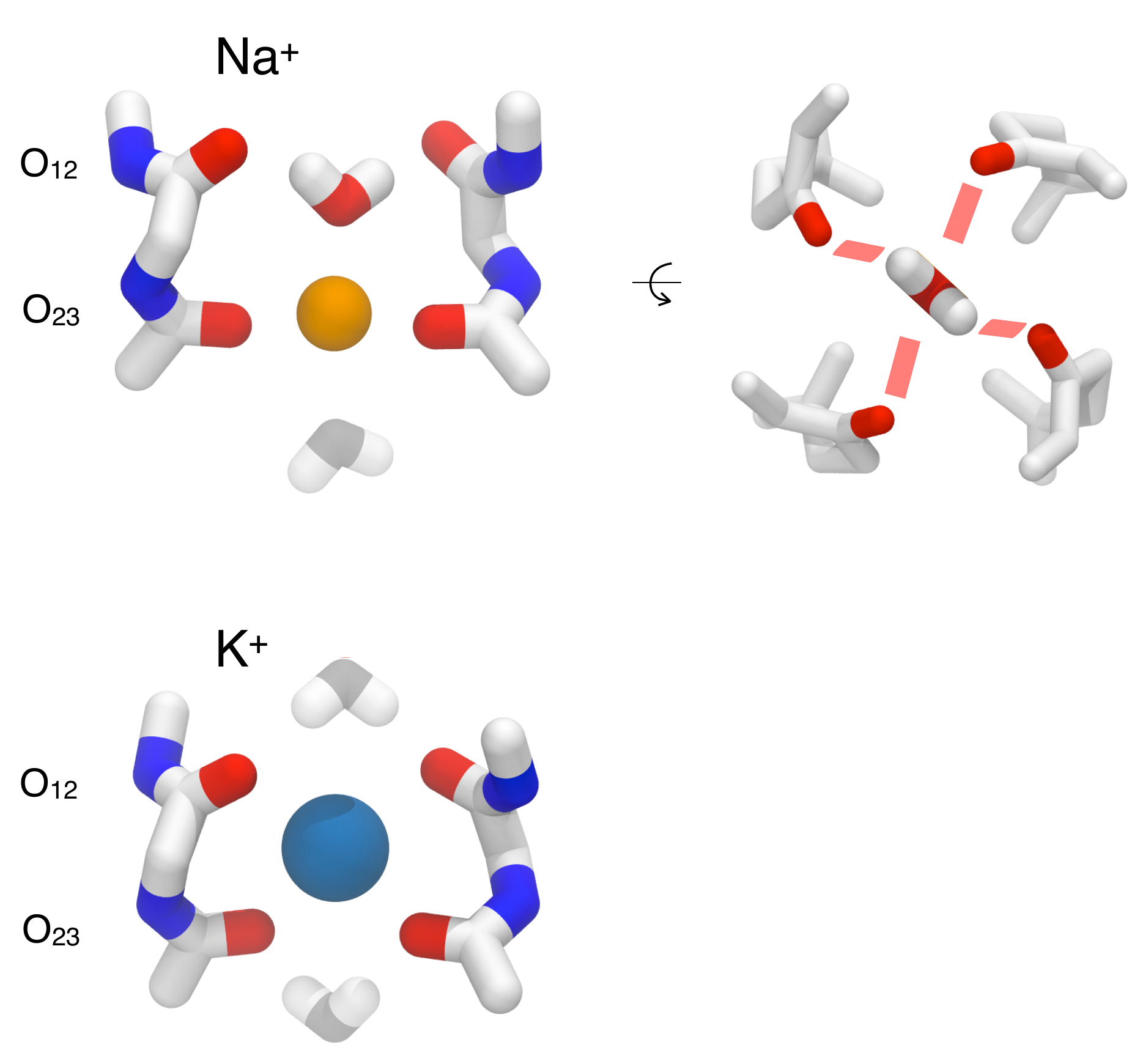}
  \end{center}
  \caption{{\bf Stable configurations inside the KcsA filter.}
    (\emph{Top}) Na$^+$ is in-plane with the $O_{23}$ carbonyls. The
    presence of a confined water molecule inside the filter stabilizes
    this position.  As shown by the 90 degrees rotated structure, this
    water molecule makes alternate hydrogen bonds (red thick lines)
    with all four $O_{12}$ carbonyls.  (\emph{Bottom}) K$^+$ is stably
    bound into the conventional $S_2$ binding site. Outside the
    filter, one and two water molecules are present for Na$^+$ and
    K$^+$, respectively (ghost waters).}
  \label{fig:spider_water}
\end{figure}

\begin{figure}[t]
  \begin{center}
    \includegraphics[width=80.0mm]{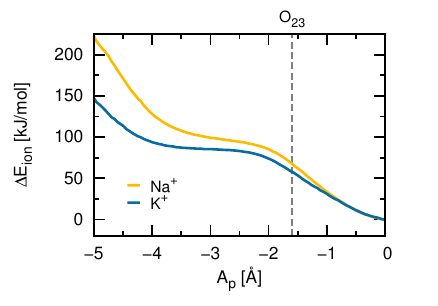}
  \end{center}
  \caption{{\bf Average interaction energy in vacuo} between the ion
    and the binding site. Data for Na$^{+}$ or K$^{+}$ is shown in
    orange and blue, respectively.  The position of the $O_{23}$
    oxygen carbonyls is shown as a vertical dashed line.}
  \label{fig:ei}
\end{figure}

This is not the case for Na$^{+}$. After few ns of simulation (see
Figure S1 in \emph{SI}), the ion hopped from $S_2$ to $O_{23}$,
assuming a configuration perfectly in plane with the 4 carbonyl
oxygens (black countoured orange area in bottom panel of
\ref{fig:histo_1D}). This position is very stable, representing the
93.6\% of the total simulation time, with a high barrier to hop back
to $S_2$ (roughly four transitions in 100 ns). Analysis of the closest
water molecules to the ion provided a mechanism for the configuration
shift.  Contrary to the case of K$^+$ where the solvent is at the
outside of the pore, one water molecule enters the channel effectively
shifting the position of Na$^+$ to $O_{23}$. This is shown by the
probability distribution of the closest water to the ion represented
as a light blue area in the figure.  A structural representation of
the confined water is illustrated in \ref{fig:spider_water}. Per se,
there is no energetic preference in shifting Na$^+$ to the $O_{23}$
position as shown by the average interaction energy $\Delta E_{ion}$
between the ion and the pore in vacuo (\ref{fig:ei}). In the
absence of water, $S_2$ is the most stable position for both K$^{+}$
and Na$^{+}$, strongly indicating that the binding site shift is due
to the presence of a confined water.

\begin{figure}[!t]
  \begin{center}
  \includegraphics[width=80.0mm]{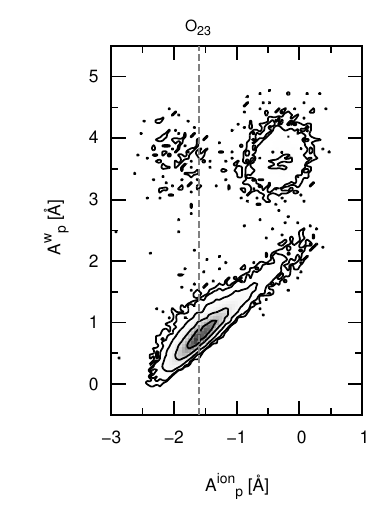}
  \end{center}
  \caption{{\bf 2D Probability density function of the ion
    ($A_p^{ion}$) and confined water ($A_p^w$) positions for the case
    of Na$^+$.} The contour lines are drawn for the probability values:
    $3\cdot 10^{-3}$, $10^{-2}$, $10^{-1}$, $1.0$, $4.0$ and $8.0$. }
  \label{fig:histo_2D}
\end{figure}

The confined water  at position $S_2$ is stabilized by multiple strong
contacts (\ref{fig:spider_water}).  The water oxygen makes
electrostatic interactions with Na$^+$, while hydrogen bonds with two of the
four carbonyls at position $O_{12}$ are formed (\ref{fig:spider_water},
top right panel).  These hydrogen bonds are also entropically stabilized, being
the confined water capable of binding to all four carbonyls of $O_{23}$ by
rotating itself around the $A_p$ axis (bond lifetime around 2 ps, see Figure~S3
in \emph{SI}).  This molecule is extremely stable, never exchanging with the
bulk. At the outside of the channel another water molecule was found at the
other side of Na$^+$, forming favorable electrostatic interactions with the ion
(ghost water in the top panel of \ref{fig:spider_water}).  The position
of the confined water correlates with the fluctuations of Na$^+$ around
$O_{23}$, indicating that the ion-water configuration behaves as a complex
(\ref{fig:histo_2D}). In the rare occasions when Na$^+$ hops back to
$S_2$, the water molecule is expelled and the complex broken ($A_p^{ion}\approx
0$ and $A_p^{w}\approx 3.8$, \ref{fig:histo_2D}).

\begin{figure}[!ht]
  \begin{center}
  \includegraphics[width=80.0mm]{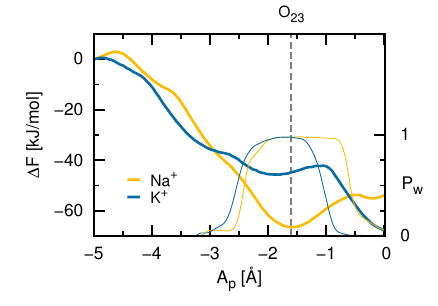}
  \end{center}
  \caption{{\bf Potential of mean force} for K$^{+}$ and Na$^{+}$
    along the axial coordinate $A_p$ (thick lines). The probability
    $P_w$ to have a water molecule inside the filter is represented as
    thin lines. Data for Na$^{+}$ or K$^{+}$ is shown in orange and
    blue, respectively.}
  \label{fig:PMF}
\end{figure}

On the other hand, for the case of K$^+$ water was only found at the outside of
the pore (ghost molecules at the bottom of \ref{fig:spider_water}).
Those waters are stabilized by the formation of hydrogen bonds with the
carbonyls of the binding site. But the interaction with the ion is much weaker
in this case, being the water oxygens facing the bulk. 

To complement the simulation study, free-energy calculations were performed. In
\ref{fig:PMF} the potential of mean force (PMF, see Methods) along the
axial coordinate $A_P$ is shown (thick lines). The most stable configurations
for Na$^+$ and K$^+$ were respectively found at positions $O_{23}$ and $S_2$,
confirming our analysis.  For both cases, position $O_{23}$ is coupled with the
presence of a water molecule inside the filter ($P_w \approx 1$ around
$O_{23}$, thin lines). Though energetically unfavorable, K$^+$ at this position
is in complex with a confined water as observed for Na$^+$.  Interestingly, the
free-energy difference between the outside of the pore ($A_p=-5$) and the most
stable binding position is remarkably similar for the two ions, indicating no
strong preference towards K$^+$ binding.  These results strongly support the
idea that the filter has the ability to bind both ions with different
mechanisms but similar strength as already suggested in
Ref.~\cite{Thompson2009Mechanism,Kim2011selective}.

\section{Discussion} It has been known, for more than a decade now, that the
selectivity filter of KcsA presents several binding sites. Each of them forms a
\emph{cage} of optimal coordination for K$^+$ by means of eight backbone
carbonyls. Most of the calculations on KcsA selectivity were based on the
relative stability of Na$^+$ over K$^+$ inside the pore with respect to the
bulk \cite{Noskov2004Control,Yu2010Two}.  That is, the solvation free-energy to
move an ion from bulk water to a specific binding site inside the selectivity
filter.  These works suggested the S2 binding site as the most selective
portion of the filter, being the free-energy difference in the $S_2$ position
much more favorable for K$^+$ compared to Na$^+$.  But this approach is not
without problems if $S_2$ is not a stable configuration for Na$^+$.  In fact,
Na$^+$ and K$^+$ might be characterized by distinct equilibrium positions
inside the pore. X-ray crystallography \cite{Thompson2009Mechanism} and
multi-ion free-energy calculations\cite{Thompson2009Mechanism,
Kim2011selective} supported this idea, showing that Na$^+$ preferentially
adopts a configuration in-plane with backbone carbonyls. The observation of
distinct binding positions has several consequences to our understanding of ion
selectivity.  Previous free-energy calculations using position restrains to the
center of the binding sites need to be extended taking into account the correct
equilibrium configurations.

Our calculations on a minimalistic model of the filter provided a mechanism for
the position shift. The presence of a confined water molecule in complex with
Na$^+$ effectively modifies the equilibrium configuration.  The confined
water is stabilized by favorable electrostatic interactions with the ion as
well as multiple hydrogen-bonds with the $O_{12}$ backbone carbonyls. The
binding position shift disappears in the absence of confined water, making the
latter an essential ingredient for the preferential position of Na$^+$. 
In-plane binding always implies the presence of confined water even
in the unfavorable case of K$^+$.

In the past, water was mostly considered in terms of solvation free-energies
and its screening effects without much attention to the molecular mechanism.
Our results reinforce the idea that  biological water is  an active player at
the molecular level \cite{ball2011}.

\section{Acknowledgments}
This work is supported by the Excellence Initiative of the German Federal and
State Governments.

\bibliography{cites}

\end{document}